\documentclass[a4paper,12pt]{article}
\usepackage{epsfig}
\usepackage[dvips,usenames]{color}
\usepackage{graphicx}

\newlength{\dinwidth}
\newlength{\dinmargin}
\begin{document}
\title{The correction of the littlest Higgs model to the Higgs
production process $e^+e^-\rightarrow e^+e^-H$ at the ILC}
\bigskip
\author{Xuelei Wang$^{a,b}$, Yaobei Liu$^{b}$, Jihong Chen$^{b}$,Hua Yang$^{c}$\\
{\small a: CCAST(World Laboratory) P.O. BOX 8730. B.J.
100080, P.R.China}\\
 {\small b: College of Physics and Information
Engineering,}\\
\small{Henan Normal University, Xinxiang, Henan, 453007,
P.R.China}
\thanks{E-mail:wangxuelei@sina.com}
\thanks{Mailing address}
\thanks{This work is supported by the National Natural Science
Foundation of China(Grant No.10375017 and 10575029).}\\
 {\small c: Department
of Mathematics and Physics, College of
Science, }\\
{\small Information Engineering University, Zhengzhou, Henan,
450001,P.R.China.}
 }
\maketitle
\begin{abstract}
\indent The littlest Higgs model is the most economical one among
various little Higgs models. In the context of the littlest Higgs
model, we study the process $e^{+}e^{-}\rightarrow e^{+}e^{-}H$ at
the ILC and calculate the correction of the littlest Higgs model
to the cross section of this process. The results show that, in
the favorable parameter spaces preferred by the electroweak
precision data, the value of the relative correction is in the
range from a few percent to tens percent. In most case, the
correction is large enough to reach the measurement precision of
the ILC. Therefore, the correction of the littlest Higgs model to
the process $e^{+}e^{-}\rightarrow e^{+}e^{-}H$ might be detected
at the ILC which will give an ideal way to test the model.
\end{abstract}
PACS number(s): 12.60Nz,14.80.Mz,12.15.Lk,14.65.Ha
\newpage
\section{Introduction}
\indent The standard model(SM) provides an excellent effective
field theory description of almost all particle physics
experiments. But the Higgs boson mass suffers from an instability
under radiative corrections in the SM. The natural argument
suggests that the cutoff scale of the SM is not much above the
electroweak scale: New physics will appear around TeV energies.
The possible new physics scenarios at the TeV scale might be
supersymmetry\cite{supersymmetry}, dynamical symmetry
breaking\cite{dynamical}, extra dimensions\cite{extra}. Recently,
a new model, known as little Higgs model, has drawn a lot of
interest and it offers a very promising solution to the hierarchy
problem in which the Higgs boson is naturally light as a result of
nonlinearly realized symmetry
\cite{little-1,little-2,little-3,littlest}. The key feature of
this model is that the Higgs boson is a pseudo-Goldstone boson of
an approximate global symmetry which is spontaneously broken by a
vev at a scale of a few TeV and thus is naturally light. The most
economical little Higgs model is the so-called
 littlest Higgs model, which is based on a $SU(5)/SO(5)$
 nonlinear sigma model \cite{littlest}. It consists of a $SU(5)$ global
 symmetry, which is spontaneously broken down to $SO(5)$ by a vacuum
 condensate $f$. In this model, a set of new heavy gauge bosons$(B_{H},Z_{H},W_{H})$ and
 a new heavy-vector-like quark(T) are introduced which just cancel
 the quadratic divergence induced by the SM gauge boson loops and the
 top quark loop, respectively. The distinguishing features of this
 model are the existence of these new particles and their
 couplings to the light Higgs. The measurement of these new particle effects might
 prove the existence of the littlest Higgs mechanism.\\

 \indent  The hunt for the Higgs and the elucidation of the symmetry
 breaking  mechanism is one of the most important goals for present and future
 high energy collider experiments. Precision electroweak measurement
  data and direct searches suggest that the Higgs boson must be relative light and its mass should
 be roughly in the range of 114.4 GeV$\sim$208 GeV at $95\%$ CL \cite{Higgs}.
 While the discovery of the Higgs at the LHC has been established for a wide range of Higgs masses, only
 rough estimates on its properties will be possible, through measurements on the couplings of the Higgs
 to the fermions and gauge boson for example \cite{LHC}.  The most precise measurements
 will be performed in the clean environment of the future high energy $e^{+}e^{-}$ linear
 collider, the International Linear Collider($ILC$) with the center of mass(c.m.) energies $\sqrt{s}$=300
 GeV-1.5 TeV \cite{ILC} and the yearly luminosity 500 $fb^{-1}$. At low energy, the main
 production processes of the Higgs boson at linear
 collider experiments are the Higgs-strahlung process
 $e^{+}e^{-}\rightarrow ZH$ and the $WW$ fusion process
 $e^{+}e^{-}\rightarrow \nu\bar{\nu}H$ and the latter is dominant
 in the large parameter space.  These two processes  have been studied in the context of the
 SM \cite{ee-SM} and the littlest Higgs model \cite{ee-LH}.
With the c.m. energy increasing, the cross section of the process
$e^{+}e^{-}\rightarrow e^{+}e^{-}H$ increases significantly. So,
at the ILC, such process becomes a welcome addition with a cross
section about $20$ fb which excesses that of $ZH$ production
around 1TeV. With the large cross section at TeV scale, the ILC
will open a promising window to probe Higgs and precisely
determinate the $ZZH$ coupling via the process
$e^{+}e^{-}\rightarrow e^{+}e^{-}H$. The calculation of the
complete electroweak correction to the process is performed in
detail in the refrence\cite{eeH}. If the Higgs boson is light, a
few percent measurement precision can be reached at the $ILC$
\cite{ILC}. The purpose of this paper is to calculate the
correction of the littlest Higgs model to the process
$e^{+}e^{-}\rightarrow e^{+}e^{-}H$ and see whether the effect on
this process can be observed at the future $ILC$ experiments.

\indent  This paper is organized as follows. In section two, we
first briefly introduce the littlest model, and then give the
production amplitude of the process. The numerical results and
discussions are presented in section three. The conclusions are
given in section four.

\section{The littlest Higgs model and the production amplitude of $e^{+}e^{-}\rightarrow e^{+}e^{-}H$
}

 \indent The littlest model is based on the a
$SU(5)/SO(5)$ nonlinear sigma model. At the scale $\Lambda_{s}\sim
4\pi$$f$, the global $SU(5)$ symmetry is broken into its subgroup
$SO(5)$ via a vacuum condensate $f$, resulting in 14 Goldstone
bosons. The effective field theory of these Goldstone bosons is
parameterized by a non-linear $\sigma$ model with gauged symmetry
$[SU(2)\times U(1)]^{2}$, spontaneously breaking down to its
diagonal subgroup $SU(2)\times U(1)$, identified as the SM
electroweak gauge group. Four of these Goldstone bosons are eaten
by the broken gauge generators, leaving 10 states that transform
under the SM gauge group as a doublet H and a triplet $\Phi$. This
breaking scenario also gives rise to four massive gauge bosons
$B_{H}$,$Z_{H}$ and $W^{\pm}_{H}$, which might produce the
characteristic signatures at the present and future high energy
collider experiments \cite{signatures-1,signatures-2,signatures-3}.\\

 \indent  After the electroweak symmetry breaking, the mass eigenstates
 are obtained via mixing between the heavy and light gauge bosons.
 They include the light (SM-like) bosons $Z_{L}$, $A_{L}$ and
$W^{\pm}_{L}$ observed at experiments, and new heavy bosons
$Z_{H}$, $B_{H}$ and $W^{\pm}_{H}$ that could be observed at
future experiments. The neutral gauge boson masses are given to
leading order by \cite{signatures-1}
\begin{eqnarray}
M^{2}_{A_{L}}&=&0,\\
M^{2}_{Z_{L}}&=&(M^{SM}_{Z})^{2}\{1-\frac{v^{2}}{f^{2}}[\frac{1}{6}+\frac{1}{4}(c^{2}-s^{2})^{2}+
\frac{5}{4}(c'^{2}-s'^{2})^{2}+\frac{\chi^2}{2}]\},\\
M^{2}_{Z_{H}}&=&(M^{SM}_{W})^{2}(\frac{f^{2}}{s^{2}c^{2}v^{2}}-1-\frac{x_{H}s^{2}_{W}}{s'^{2}c'^{2}c^{2}_{W}}),\\
M^{2}_{B_{H}}&=&(M^{SM}_{Z})^{2}s^{2}_{W}(\frac{f^{2}}{5s'^{2}c'^{2}v^{2}}-1+\frac{x_{H}c^{2}_{W}}{4s^{2}c^{2}s^{2}_{W}}),
\end{eqnarray}
with $\chi=\frac{4fv'}{v^{2}},~
x_{H}=\frac{5}{2}gg'\frac{scs'c'(c^{2}s'^{2}+s^{2}c'^{2})}{5g^{2}s'^{2}c'^{2}-g's^{2}c^{2}}$,
where~$v$=246 GeV is the elecroweak scale, $v'$ is the vacuum
expectation value of the scalar $SU(2)_{L}$ triplet and
$s_{W}(c_{W})$ represents the sine(cosine) of the weak mixing
angle. The parameter $\chi<1$ parametrizes the ratio of the triple
and doublet vacuum expectation values. In the following
calculation, we will take $\chi$=0.5. \\

\indent  Taking account of the gauge invariance of the Yukawa
coupling and the $U(1)$ anomaly cancellation, we can write the
couplings of the neutral gauge bosons $V_i(V_i=Z_{L},B_{H},Z_{H})$
to a pair of electrons in the form
$\wedge_{\mu}^{V_i\bar{e}e}=i\gamma_{\mu}(g^{V_i\bar{e}e}_{V}+g^{V_i\bar{e}e}_{A}\gamma^{5})$
and denote the couplings of two gauge bosons to Higgs as
$\wedge_{\mu\nu}^{HV_iV_j}$. $g^{V_i\bar{e}e}_V,g^{V_i\bar{e}e}_A$
and $\wedge_{\mu\nu}^{HV_iV_j}$ can be written as
\cite{signatures-1}:
\begin{eqnarray}
g^{Z_{L}\bar{e}e}_{V}&=&-\frac{e}{4s_{W}c_{W}}\{(-1+4s^{2}_{W})-\frac{v^{2}}
{f^{2}}[\frac{1}{2}c^{2}(c^{2}-s^{2})-\frac{15}{2}(c'^{2}-s'^{2})(c'^{2}-\frac{2}{5})]\},\\
g^{Z_{L}\bar{e}e}_{A}&=&-\frac{e}{4s_{W}c_{W}}\{1+\frac
{v^{2}}{f^{2}}[\frac{1}{2}c^{2}(c^{2}-s^{2})+\frac{5}{2}(c'^{2}-s'^{2})(c'^{2}-\frac{2}{5})]\},\\
g^{Z_{H}\bar{e}e}_{V}&=&-\frac{ec}{4s_{W}s},
 \hspace{6cm}g^{Z_{H}\bar{e}e}_{A}=\frac{ec}{4s_{W}s},\\
g^{B_{H}\bar{e}e}_{V}&=&\frac{e}{2c_{W}s'c'}(\frac{3}{2}c'^{2}-\frac{3}{5}),
 \hspace{4cm}g^{B_{H}\bar{e}e}_{A}=\frac{e}{2c_{W}s'c'}(\frac{1}{2}c'^{2}-\frac{1}{5}),\\
\wedge_{\mu\nu}^{HZ_{L}Z_{L}}&=&\frac{ie^{2}vg_{\mu\nu}}{2s^{2}_{W}c^{2}_{W}}\{1-\frac{v^{2}}
{f^{2}}[\frac{1}{3}-\frac{3}{4}\chi^{2}+\frac{1}{2}(c^{2}-s^{2})^{2}+\frac{5}
{2}(c'^{2}-s'^{2})^{2}]\},\\
\wedge_{\mu\nu}^{HZ_{H}Z_{H}}&=&-\frac{ie^{2}}{2s_{W}^{2}}vg_{\mu\nu},
\hspace{4.2cm}\wedge_{\mu\nu}^{HB_{H}B_{H}}=-\frac{ie^{2}}{2c_{W}^{2}}vg_{\mu\nu},\\
\wedge_{\mu\nu}^{HZ_{L}Z_{H}}&=&-\frac{ie^{2}(c^{2}-s^{2})vg_{\mu\nu}}{4s^{2}_{W}c_{W}sc},
 \hspace{2.7cm}\wedge_{\mu\nu}^{HZ_{L}B_{H}}=-\frac{ie^{2}(c'^{2}-s'^{2})vg_{\mu\nu}}{4s_{W}c^{2}_{W}s'c'},\\
\wedge_{\mu\nu}^{HZ_{H}B_{H}}&=&-\frac{ie^{2}vg_{\mu\nu}}{4s_{W}c_{W}}\frac{(c^{2}s'^{2}+s^{2}c'^{2})}{scs'c'}.
\end{eqnarray}

 \indent  The tree-level $e^{+}e^{-}\rightarrow e^{+}e^{-}H$ process is built up from
 the $s$-channel diagrams originating from $e^{+}e^{-}\rightarrow HV_i$ and
 the $t$-channel diagrams which are the fusion types.
 The relevant tree level Feynman diagrams of the process are shown in Fig.1.
\begin{figure}[t]
\begin{center}
\epsfig{file=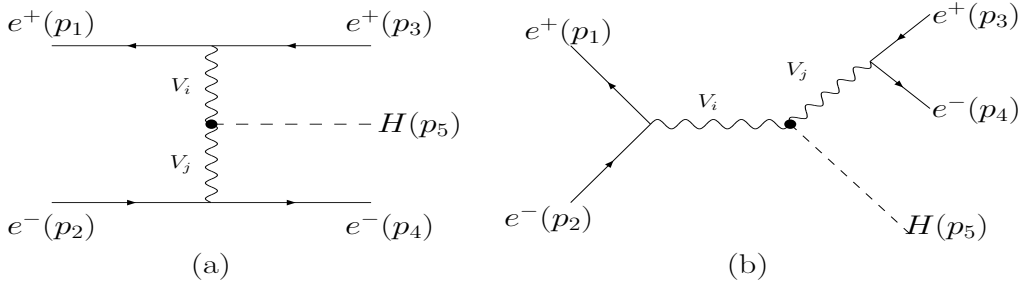,width=450pt,height=500pt} \vspace{-12cm}
\caption{\small Feynman diagrams of the process
$e^{+}e^{-}\rightarrow e^{+}e^{-}H$ in the littlest Higgs model.}
\label{fig1}
\end{center}
\end{figure}

 The invariant production amplitudes of the process
 can be written as:
\begin{equation}
 M=-\sum_{V_{i,j}=Z_{L},Z_{H},B_{H}}M_{a}^{V_{i}V_{j}}+
 \sum_{V_{i,j}=Z_{L},Z_{H},B_{H}}M_{b}^{V_{i}V_{j}},
 \end{equation}
 with
 \begin{eqnarray*}
 M_{a}^{V_{i}V_{j}}&=&\overline{u}_{e}(p_{4})\wedge_{\mu}^{V_{j}\bar{e}e}u_{e}(p_{2})
G^{\mu\rho}(p_{4}-p_{2}, M_{V_{j}})
 \wedge_{\rho\tau}^{HV_{i}V_{j}}
G^{\tau\nu}(p_{1}-p_{3}, M_{V_{i}})
 \overline{v}_{e}(p_{1})\wedge_{\nu}^{V_{i}\bar{e}e}v_{e}(p_{3}),\\
 M_{b}^{V_{i}V_{j}}&=&\overline{u}_{e}(p_{4})\wedge^{V_{j}\bar{e}e}_{\mu}v_{e}(p_{3})
G^{\mu\rho}(p_{3}+p_{4}, M_{V_{j}})
\wedge_{\rho\tau}^{HV_{i}V_{j}} G^{\tau\nu}(p_{1}+p_{2},
M_{V_{i}})
 \overline{v}_{e}(p_{1})\wedge_{\nu}^{V_{i}\bar{e}e}u_{e}(p_{2}).
 \end{eqnarray*}
 Here, $G^{\mu\nu}(p,M)=\frac{-ig^{\mu\nu}}{p^{2}-M^{2}}$ is the propagator of the
 particle. There is minus sign difference in the contributions of
 s-channel and t-channel diagrams.
 We can see that the one source of the corrections of the littlest Higgs model to the process
 arises from the new heavy gauge bosons $Z_H,B_H$. On
 the other hand, the
 littlest Higgs model can generate the correction to the mass of $Z$ boson in the SM
and to the tree-level coupling vertices, which can also produce
the correction to the process. In our numerical calculation, we
will also take into account such correction effect. It should be
noted that the masses of gauge bosons vary with the parameters $c$
and $c'$, and $M_{Z_H}^2$ can equal to $e^+e^-$ c.m. energy square
$(p_1+p_2)^2$ for the certain values of parameters which will
cause the s-channel resonance effect in Fig.1(b). For the gauge
boson propagators of Fig.1(b) connecting outgoing $e^+e^-$, the
time-like momentum can also hit the light gauge boson pole which
can also cause some resonance effect. In this case, we should take
into account the effect of the widths of gauge bosons in the
calculation. i.e., we should take the complex mass term
$M_{V_{i}}^2-iM_{V_i}\Gamma_{V_i}$ instead of the simple gauge
boson mass term $M_{V_{i}}^2$ in the gauge boson propagators. The
$-iM_{V_i}\Gamma_{V_i}$ term is important in the vicinity of the
resonance. We can take $\Gamma_{Z_L}=2.4952$ GeV( the total
experimental width of observed Z boson). The main decay modes of
$B_H$ and $Z_H$ are $V_i\rightarrow f\bar{f}$(f represents any
quarks and leptons in the SM) and $V_i\rightarrow ZH$. The decay
widths of these modes have been explicitly given in references
\cite{signatures-1,yue}.

\indent With above production amplitudes, we can obtain the
production cross section directly. In the calculation of the cross
section, instead of calculating the square of the amplitudes
analytically, we calculate the amplitudes numerically by using the
method of the references\cite{HZ} which can greatly simplify our
calculation.

\section{ The numerical results and discussions}

The process $e^+e^-\rightarrow e^+e^-H$ has been studied in the SM
and the one-loop electroweak correction has been
considered\cite{eeH}. Because the t-channel contribution to the
cross section rises depending on $Log(s/M_{V_i})$, the total cross
section can reach the order of $10 $ fb with $\sqrt{s}=800$
 GeV. The electroweak correction is negative and in the range from $-2\%$ to
$-4\%$. In this paper, we calculate the littlest Higgs correction
to the process in the tree level.

In the numerical calculation, we take the input parameters as
 $M_{Z}^{SM}=91.187$ GeV, $s_{W}^{2}=0.2315$ \cite{data}.  For the light Higgs boson H, in this paper,
 we only take the illustrative value $M_{H}=120$ GeV. The c.m. energy of the ILC is assumed
as $\sqrt{s}$=800 GeV. In the littlest Higgs model, there are
three free parameters, $f,c,c'$, involved in the production
amplitude. The custodial $SU(2)$ global symmetry is
 explicitly broken, which can generate large contributions to the
 electroweak observables. However, if we carefully adjust the $U(1)$ section of the theory
 the contributions to the electroweak observables can be reduced and
 the constraints become
 relaxed. The scale parameter $f=1\sim2$ TeV is allowed for the mixing
 parameters $c$ and $c^{'}$ in the ranges of
 $0 \sim 0.5,0.62 \sim 0.73$ \cite{constraints}. Taking into account the constraints on $f,c,c'$, we
 take them as the free parameters in our numerical calculation.
 The numerical results are summarized in Figs.(2-4)

\begin{figure}[h]
\begin{center}
\scalebox{0.75}{\epsfig{file=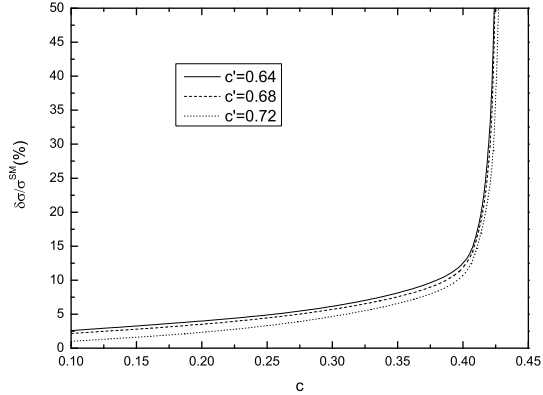}}\\
\end{center}
\caption{\small The relative correction $\delta\sigma/\sigma^{SM}$
as a function of the mixing parameter c for f=1 TeV, $M_{H}=120$ GeV
and three values of the mixing $c'$.}
\end{figure}

\indent  The relative correction $\delta\sigma/\sigma^{SM}$ is
plotted in Fig.2 as a function of the mixing parameter c for $f$=1
TeV, $c^{'}=0.64, 0.68, 0.72$ and $M_{H}=120$ GeV, in which
$\delta\sigma=\sigma^{tot}-\sigma^{SM}$ and $\sigma^{SM}$ is the
tree-level cross section of $e^{+}e^{-}H$ production predicted by
the SM. From Fig.2, we can see that the relative correction
$\delta\sigma/\sigma^{SM}$ increases sharply when c approaches
0.45. This is because in this range the mass of $Z_H$ may equal to
the $c.m.$ energy $\sqrt{s}$(800 GeV) which can make the large s
-channel resonance effect in Fig.1(b). The value of the relative
correction varies in a wide range from a few percent to tens
percent. There exists a special case that when $c'=\sqrt{2/5}$,
the heavy photon $B_{H}$ has no contribution to the process
because the coupling of the $B_{H}$ to the electrons vanishs. \\

\begin{figure}[b]
\begin{center}
\scalebox{0.8}{\epsfig{file=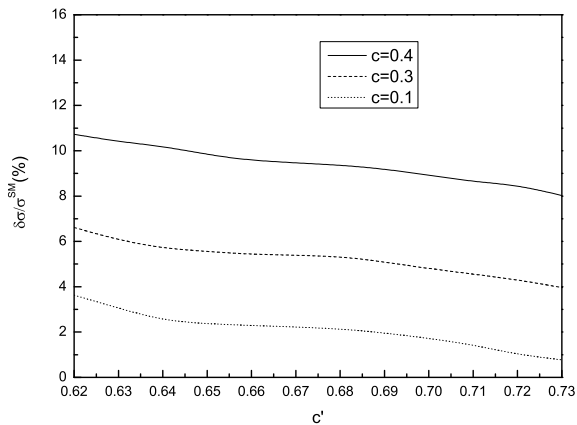}}
\end{center}
\caption{\small The relative correction $\delta\sigma/\sigma^{SM}$
as a function of the mixing parameter $c^{'}$ for f=1 TeV,
$M_{H}=120$ GeV and c=0.1(dotted line), 0.3(dashed line) and
0.4(solid line).}
\end{figure}

 \indent To see the dependence of the relative correction on the parameter $c'$, in Fig.3,
we plot $\delta\sigma/\sigma^{SM}$ as a function of the mixing
parameter $c'$ for f=1 TeV, $M_{H}=120$ GeV, and three values of
the mixing parameter c. We can see that the relative correction
decrease with the mixing parameter c' increasing and is more
sensitive to the parameter c. For $c>0.3$, the value of
$\delta\sigma/\sigma^{SM}$ is larger than $5\%$, which might be
detected in the future LC experiments.

In general, the contributions of the littlest Higgs model to the
observables are dependent on the factor $1/f^{2}$. To see the
effects of f and the Higgs mass on the cross section, in Fig 4, we
plot $\delta\sigma/\sigma^{SM}$ as a function of $f$ for three
values of Higgs mass($M_H=120,150,180$ GeV) and take $c=0.4$,
$c^{'}=0.68$. One can see that the relative correction drops
sharply with f increasing. For example, the relative correction is
below $2\%$ when $f=2$ TeV. This case is similar to the
contributions of the littlest Higgs model to other observables. On
the other hand, the curves show that the relative correction is
not sensitive to the Higgs boson mass.

\begin{figure}[ht]
\begin{tabular}{cc}
~~~~~~~~~~~~~~~~~~~~~~~\scalebox{0.85}{\epsfig{file=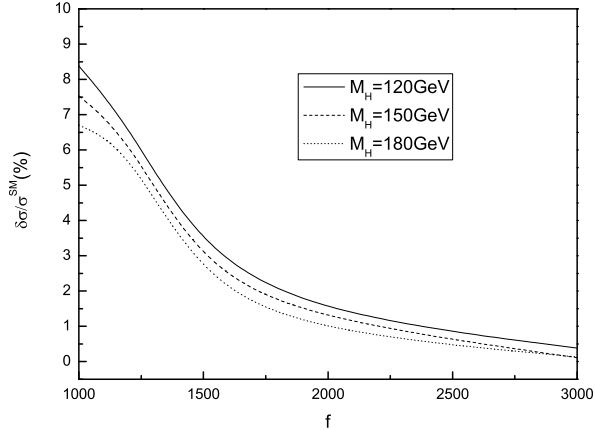}}
\end{tabular}
\caption{\small The relative correction $\delta\sigma/\sigma^{SM}$
as a function of the the scale parameter f for $c=0.4$, $c'=0.68$
and three values of the Higgs mass.}
\end{figure}

In contrast to the electroweak correction, the correction of the
littlest Higgs model is positive. Therefore, such significant
positive correction is a definite signal of the new physics model.
As has been mentioned above, the total cross section of
$e^{+}e^{-}\rightarrow e^{+}e^{-}H$ can reach the order of 20 fb at
the ILC. This cross section amounts to about $10^4$ events with the
integrated luminosity of 1000 $fb^{-1}$. The $1\sigma$ statistical
error corresponds to about $1\%$ precision. Even we consider the
systemic error of the ILC, the ILC can measure the cross section
with a few percent precision\cite{ILC} and the relative correction
of the littlest Higgs model to the cross section is comparable to
the ILC measurement precision. So, such correction might be detected
at the ILC.

\section{Conclusion}
 \indent The little Higgs model, which
can solve the hierarchy problem, is a promising alternative model of
new physics beyond the standard model. Among the various little
Higgs models, the littlest Higgs model is one of the simplest and
phenomenologically viable models. The distinguishing feature of this
model is the existence of the new scalars, the new gauge bosons, and
the vector-like top quark. These new particles contribute to the
experimental observables which could provide some clues of the
existence of the littlest Higgs model. In this paper, we study the
potential to detect the contribution of the littlest Higgs model via
the process $e^+e^-\rightarrow e^+e^-H$ at the future $ILC$
experiments.

\indent In the parameter spaces($f=1\sim2$ TeV, $c=0\sim0.5$,
$c'=0.62\sim0.73$) limited by the electroweak precision data, we
calculate the correction of the littlest Higgs model to the cross
section of the process $e^{+}e^{-}\rightarrow e^{+}e^{-}H$. We
find that the correction is significant even when we consider the
constraint of electroweak precision data on the parameters. The
contribution mainly comes from $t$-channel due to the high c.m.
energy enhancing effect. But when $c$ approaches 0.45, the mass of
heavy $Z_H$ can equal to the $e^+e^-$ c.m. energy which can cause
the resonance effect arising from s-channel $Z_H$ propagator, and
the correction sharply increases in this case. The relative
correction varies from a few percent to tens percent. The littlest
Higgs model is a weak interaction theory and it is hard to detect
its contributions and measure its couplings at the LHC. With the
high c.m. energy and luminosity, the future ILC will open an ideal
window to probe into the littlest Higgs model and study its
properties. In most case, the relative correction of the littlest
model to the process $e^{+}e^{-}\rightarrow e^{+}e^{-}H$ is large
enough for people to measure the contribution of the model with
the high precision at the ILC. Therefore, the process
$e^{+}e^{-}\rightarrow e^{+}e^{-}H$ will open an ideal window to
test the littlest Higgs model.

\end{document}